\documentclass[conference,10pt]{IEEEtran}
\usepackage{epsfig,epsf,rotating,setspace,latexsym,amsmath,amssymb,amsfonts,bm,theorem,subfigure,epstopdf,cite,authblk,bbm}
\usepackage{algorithm}
\usepackage[noend]{algpseudocode}
\usepackage{color}
\usepackage{mathtools}
\usepackage{multirow}
\usepackage{soul}

\algrenewcommand\algorithmicforall{\textbf{foreach}}
\algrenewcommand\algorithmicindent{.8em}
\DeclarePairedDelimiter{\ceil}{\lceil}{\rceil}

\newtheorem{theorem}{Theorem}

\newtheorem{remark}{Remark}

\IEEEoverridecommandlockouts
\allowdisplaybreaks

\begin{document}

\title{Distributed Optimization with Feasible Set Privacy}
 
\author{Shreya Meel \qquad Sennur Ulukus\\
        \normalsize Department of Electrical and Computer Engineering\\
        \normalsize University of Maryland, College Park, MD 20742\\
        \normalsize  \emph{smeel@umd.edu} \qquad \emph{ulukus@umd.edu}}

\maketitle

\begin{abstract}
We consider the setup of a constrained optimization problem with two agents $E_1$ and $E_2$ who jointly wish to learn the optimal solution set while keeping their feasible sets $\mathcal{P}_1$ and $\mathcal{P}_2$ private from each other. The objective function $f$ is globally known and each feasible set is a collection of points from a global alphabet. We adopt a sequential symmetric private information retrieval (SPIR) framework where one of the agents (say $E_1$) privately checks in $\mathcal{P}_2$, the presence of candidate solutions of the problem constrained to $\mathcal{P}_1$ only, while learning no further information on $\mathcal{P}_2$ than the solution alone. Further, we extract an information theoretically private threshold PSI (ThPSI) protocol from our scheme and characterize its download cost. We show that, compared to privately acquiring the feasible set $\mathcal{P}_1\cap \mathcal{P}_2$ using an SPIR-based private set intersection (PSI) protocol, and finding the optimum, our scheme is better as it incurs less information leakage and less download cost than the former. Over all possible uniform mappings of $f$ to a fixed range of values, our scheme outperforms the former with a high probability. 
\end{abstract}

\section{Introduction}\label{intro}

\begin{figure}[t]
    \centering
    \includegraphics[width=0.4\textwidth]{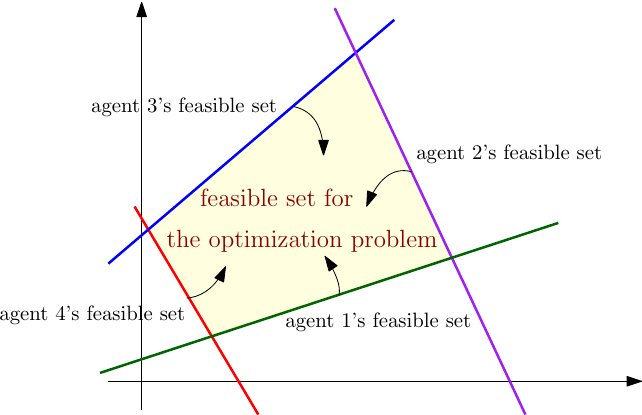}
    \centering
    \caption{Feasible set for a distributed optimization problem with four agents each with a linear inequality constraint.}
    \label{visualize}
\end{figure}

In distributed optimization, agents collaborate to find the optimal solution of a global objective function. Each agent has its own feasible set and thus the solution of the optimization problem should be present in the intersection of all feasible sets as shown in Fig.~\ref{visualize}. The feasible set contains sensitive information of an agent, and should be kept private. For instance, consider the problem of allocating charging schedules to electric vehicles (EVs) \cite{pappasDCOP}, to minimize load fluctuations on a power grid subject to maximum charge and energy constraints for each EV. Then, the maximum charge for an EV being zero suggests that the owner of the EV (agent) is away, leaking private information about the agent. Most importantly, the iterative optimization algorithms \cite{asumanDCOP} involve exchanging solution estimates among agents over iterations which, on collusion of all but one agent reveals information on the target agent's feasible set. 

Differential privacy (DP) \cite{dworkDP} based optimization algorithms are a standard solution to alleviate this problem to some extent by adding noise to the estimates before exchanging them. Besides feasible sets, the local functions whose sum is the global objective also reveal sensitive agent information, and guaranteeing DP to agents in these scenarios was studied in \cite{vaidyaPDOP, diggaviDCOP, vaidyaDCOP}. However, the DP-based approaches compromise the accuracy of solutions owing to the noise added, while failing to guarantee information theoretic privacy.

We address this research gap in the simplest case by formulating a two-agent optimization problem, where the feasible set of each agent is a finite list of values from a universal alphabet and the function $f$ is known to both agents. To search for the optimum set $\mathcal{P}^*$, one of the agents \emph{privately} learns whether the best solutions in its own feasible set are also present in the feasible set of the other agent, while learning \emph{no further information} on the feasible set of the latter. The agents achieve this by formulating the feasible sets as incidence vectors \cite{zhushengPSI, zhushengMultiPSI} and one of them checks the cardinality of elements achieving the best function value appearing in the other agent by invoking the cardinality PSI (CarPSI) protocol, which privately determines the cardinality of the intersection of two sets. If the returned value is zero, this means that none of those candidate elements are present in the intersection. The agent re-invokes CarPSI, this time with the next best set of elements. Once the returned value is positive, the search for the optimum function value is complete. 

Next, the agent finds $\mathcal{P}^*$ by executing FindPSI method on the subset given the knowledge of its cardinality. The above procedure can be viewed as a sequential application of a two-user threshold private set intersection (ThPSI) protocol-wherein the set intersection is evaluated only if its cardinality is greater than or equal to a threshold, while revealing no further information about the sets as in PSI \cite{tsudikPSI}. An important instance of ThPSI arises in a ride sharing app where each passenger wants to share a ride only when there is a sufficient overlap between their respective routes. In the context of our scheme, the ThPSI operates on subsets dictated by the function values and the threshold is 1 in our case. Hence, we obtain a ThPSI protocol guaranteeing information theoretic privacy unlike the existing protocols of \cite{zhangthPSI, ghoshthPSI} where a security parameter regulates information leakage.

We adopt a client-server framework as in the information theoretic formulation of private information retrieval (PIR) \cite{jafarPIR, ulukusPIRLC}, with the added requirement of database security to build symmetric PIR (SPIR) \cite{jafarSPIR} based methods CarPSI and FindPSI. They are executed by one of the agents, who we call the client. We quantify the information leaked to the client by the mutual information between $\mathcal{P}^*$ and feasible set of the server. We measure the download cost of our scheme by the maximum number of symbols the client receives from the server to find $\mathcal{P}^*$. The download cost of our scheme varies from a minimum of $2$ to a maximum equal to that of PSI \cite{zhushengPSI} under the same client-server allocation. This shows that the client incurs greater download cost for most functions if he learns the entire feasible set first and then evaluates $f$ on it to learn $\mathcal{P}^*$. Specifically, under uniform random mappings of $f$, the probability that both download costs are equal is low, as shown through our numerical result. Further, unlike our scheme, by learning the set intersection, the client learns more information on the server's feasible set than the minimum. 

\section{System Model}\label{mod}

In our model, there is a finite indexed alphabet set $\mathcal{P}_{alph}$ of size $K=|\mathcal{P}_{alph}|$ and two entities (agents) $E_1$ and $E_2$. Entity $E_i, i=1,2$ has a feasible set $\mathcal{P}_i , |\mathcal{P}_i|=P_i$ as a subset of $\mathcal{P}_{alph}$. We cast each $\mathcal{P}_i$ into a $K$-length binary vector $X_i$ which indicates the presence or absence of an item by $1$ and $0$, respectively. We call $X_i$ the \emph{incidence vector} of $E_i$. For $i=1,2$, $X_i$ is a column vector of dimension $K$ with,
\begin{align}
     X_i(k) =\begin{cases} 
      1, & \text{if }\mathcal{P}_{alph}(k) \in \mathcal{P}_i, \\
      0, & \text{otherwise}, 
   \end{cases} 
   &&  \forall k\in[K].
\end{align}
Let $\mathcal{I}_1$ and $\mathcal{I}_2$ represent the set of indices having $1$ in vectors $X_1$ and $X_2$, respectively. That is, $\mathcal{P}_{alph}(\mathcal{I}_i)=\mathcal{P}_i$, $i=1,2$. Each entity $E_i$ is equipped with $N_i$ non-colluding databases, where the vector $X_i$ of $K$ bits is replicated and stored. The numbers $P_i$ and $N_i$, $i= 1, 2$ are known to both entities.
 
The goal of $E_1$ and $E_2$ is to optimize a function $f$ over the joint feasible set $\mathcal{P}_1\cap \mathcal{P}_2$. For a minimization problem with $f$ as the objective function, they should find the solution to
\begin{align}
    &\underset{x}{\text{minimize}}
    && f(x) \nonumber\\
    &\text{subject to}
    && x\in \mathcal{P}_1\cap \mathcal{P}_2
\end{align}
where $\mathcal{P}_1\cap \mathcal{P}_2$ is non-empty. To ensure non-emptiness of the set intersection, it is sufficient to have $K<P_1+P_2$. Since we are interested in the worst case costs, we model the function $f$ as a uniform random mapping,
\begin{align}
    f:\mathcal{P}_{alph}\longmapsto \mathcal{T}
\end{align}
independent of the realizations of $\mathcal{P}_1$ and $\mathcal{P}_2$. The function assumes $T$ distinct values, as represented by the set $\mathcal{T}$, $|\mathcal{T}|=T$ where each element in $\mathcal{P}_{alph}$ is uniformly and independently assigned one of the values in $\mathcal{T}$. Thus, for any $x\in \mathcal{P}_{alph}$ and  $y \in \mathcal{T}$, $\mathbb{P}(f(x)=y)=\frac{1}{T}$.

The entities are honest but curious, in the sense that each of them follows the protocol truthfully, but is eager to learn about the private feasible set of the other. The protocol requires one of the entities to initiate the communication. This entity is the \emph{client} while the other entity is the \emph{server}. By the end of the protocol, the client learns the optimal solution first, and later conveys it directly to the server. Without loss of generality, let $E_1$ be the client and $E_2$ be the server. Before starting any communication between the entities, the $N_2$ databases of the server $E_2$ share a set of common randomness symbols $\mathcal{S}=\{S_1,S_2,\ldots, S_{m}\}$ where $m=\ceil{\frac{P_1}{N_2-1}}$, independent of $f$, $\mathcal{P}_1$ and $\mathcal{P}_2$. The databases do not collude hereafter. One approach to solve this problem is: $E_1$ first finds the joint feasible set $\mathcal{P}_1\cap \mathcal{P}_2$ privately, following the existing PSI protocol, and then evaluates the function at the values in the feasible set only, selects the best ones among them to obtain the solution. We refer to this as the \emph{naive approach}. Here, the entire $\mathcal{P}_1\cap \mathcal{P}_2$ is leaked to $E_1$. Thus, the values of $X_2$ at indices in $\mathcal{I}_1$ are revealed to $E_1$, and the information $I(\mathcal{P}_1\cap \mathcal{P}_2;\mathcal{P}_2|\mathcal{P}_1,f)$ thus leaked to $E_1$ is not minimum.

To discourage this, we restrict the amount of information on $\mathcal{P}_2$ leaked to $E_1$ during the optimization to the minimum value. This amount of information leakage is inevitable, irrespective of the scheme adopted and is less than the information revealed to client on learning the entire feasible set. We term this as the \emph{nominal information leakage} and is defined as the amount of information on $\mathcal{P}_2$ leaked to the client $E_1$ from learning the optimum solution set alone. Let $\mathcal{P}^*\subset \mathcal{P}_1 \cap \mathcal{P}_2$ denote the optimal solution set. $\mathcal{P}^*$ can be completely determined with the knowledge of $f$, and $\mathcal{P}_1\cap \mathcal{P}_2$, i.e.,
\begin{align}
    H(\mathcal{P}^*|\mathcal{P}_1\cap\mathcal{P}_2,f)=0.
\end{align}

First, $E_1$ starts the protocol by designing and sending a query $Q_n, n\in [N_2]$ to each database $n$ of $E_2$. The queries are generated without the knowledge of $E_2$'s feasible set $\mathcal{P}_2$, i.e.,
\begin{align}
    I(Q_{1:N_2};\mathcal{P}_2)=0.
\end{align}
With the received query, each database $n$ of $E_2$ responds to $E_1$ with an answer $A_n$, which is a deterministic function of the query $Q_n$, set $\mathcal{P}_2$ (or vector $X_2$) and $\mathcal{S}$,
\begin{align}
    H(A_n|Q_n,\mathcal{P}_2,\mathcal{S})=0, && n\in[N_2].
\end{align}
Using the sent queries $Q_{1:N_2}$, the collected answers $A_{1:N_2}$ and its feasible set $\mathcal{P}_1$, client $E_1$ exactly finds the optimal solution set $\mathcal{P}^*$. This is represented by the reliability constraint,
\begin{align}
\text{[reliability]} && H(\mathcal{P}^*|Q_{1:N_2}, A_{1:N_2},\mathcal{P}_1,f)=0.
\end{align}
To ensure the privacy of $E_1$, no information on its feasible set $\mathcal{P}_1$ should be revealed to any individual database of $E_2$ by the query it receives and the answer it generates, 
\begin{align}
\hbox{[$E_1$ privacy]} && I(\mathcal{P}_1;Q_n, A_n, \mathcal{S}|\mathcal{P}_2,f)=0, && n\in[N_2].
\end{align}
For $E_2$'s privacy, no information on $\mathcal{P}_2$ should be revealed collectively by the queries and answers from the $N_2$ databases to $E_1$ beyond the nominal information leakage,
\begin{align}\label{nominfoleak}
 \text{[$E_2$ privacy]} && I(\mathcal{P}_2;Q_{1:N_2}, A_{1:N_2}|\mathcal{P}_1,f)=I(\mathcal{P}^*;\mathcal{P}_2|\mathcal{P}_1,f).
\end{align}
Since $\mathcal{P}^*$ is a subset of $\mathcal{P}_1 \cap \mathcal{P}_2$, \eqref{nominfoleak} is less than or equal to $I(\mathcal{P}_1\cap \mathcal{P}_2;\mathcal{P}_2|\mathcal{P}_1,f)$.

The download cost, $D$ is the maximum number of symbols (of the field) that $E_1$ downloads from $E_2$ till $\mathcal{P}^*$ is found. Given a fixed realization of feasible sets $\mathcal{P}_1$ and $\mathcal{P}_2$, $D$ depends on the realization of $f$. In the naive approach, the download cost is fixed to $D_{PSI}$ irrespective of $f$ as stated in Theorem~\ref{thrm1}.

\begin{theorem}\label{thrm1}
    The download cost of the naive approach is equal to that of PSI \cite{zhushengPSI}, and is given by,
    \begin{align}
    D_{PSI}=D_{PSI}(P_1,N_2)=\bigg\lceil \frac{P_1N_2}{N_2-1}\bigg\rceil.
\end{align}
\end{theorem}

\section{Main Results} \label{res}

The equi-cost elements, i.e., those with equal function values in $\mathcal{P}_1$ are ordered from best to worst and let $\mathcal{J}_r$ be the set of indices in $X_1$ corresponding to the $r$th best solution in $\mathcal{P}_1$, with  $f_{i_r}$ as the respective function value. Let $\bm{\alpha}=[\alpha_1, \alpha_2, \ldots, \alpha_L]$ be a vector with $\alpha_r$ denoting the multiplicity of the $r$th best solution, $r\in[L]$. Each entry $\alpha_r\geq 1$ and $\sum_{r=1}^L\alpha_r = P_1$. Note that $|\mathcal{J}_r|=\alpha_r$. We define the $K$-length column vector $X_{\mathcal{J}_r}$ as follows,
\begin{align}
     X_{\mathcal{J}_r}=\sum_{j\in\mathcal{J}_r }{e}_j,
\end{align}
where ${e}_j$ is a $K$-length column vector with zeros everywhere except the $j$th position.

\begin{theorem}\label{thm1}
     For a 2-agent optimization problem, given the feasible sets $\mathcal{P}_1$ and $\mathcal{P}_2$, the download cost $D$ depends on the index $R$ of the optimal function value $f_{i_R}$ in $\bm{\alpha}$ and is given by,
     \begin{align}\label{dlcost1}
    D&=
    \begin{cases}
     \big\lceil\frac{(R+\alpha_R-1)N_2}{N_2-1}\big\rceil, &   \text {if }\alpha_R > X_{\mathcal{J}_R}^TX_2, \\
     \big\lceil\frac{RN_2}{N_2-1}\big\rceil, & \text{if } \alpha_R = X_{\mathcal{J}_R}^TX_2.
    \end{cases}
\end{align}
\end{theorem}

\begin{remark}
    If $\alpha_R+R=P_1+1$, and the returned cardinality $X_{\mathcal{J}_R}^TX_2<\alpha_R$, then the download cost is maximum, given by $\big\lceil\frac{P_1N_2}{N_2-1}\big\rceil$ which is equal to that incurred by the naive approach. On the other extreme, if $R=1$ and $\alpha_R=1$, then the download cost is minimum, given by $\big\lceil\frac{N_2}{N_2-1}\big\rceil=2$.
\end{remark}

\begin{remark}\label{remPeq}
Given a fixed realization of feasible sets $\mathcal{P}_1$ and $\mathcal{P}_2$, let $P_{eq}$ denote the probability under the function space, when $D$ and $D_{PSI}$ are equal. We calculate $P_{eq}$ and show through numerical results that this probability is small.
\end{remark}

\begin{theorem}\label{thPSI}
    For a ThPSI problem with threshold $t$, the worst case download cost $D_{ThPSI}$ is,
    \begin{align}
    D_{ThPSI}=
        \begin{cases}
           2, & \text{if } M<t, \text{ or } M=P_1, \\
           D_{PSI}, & \text{if } t\leq M\leq P_1-1.
        \end{cases}
    \end{align}
\end{theorem}

\begin{figure}[t]
   \centering
   \includegraphics[width=0.3\textwidth]{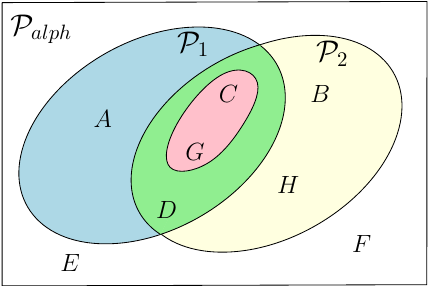}
   \caption{Illustration of Example~1 with $\mathcal{P}^*=\{C,G\}$.}
   \label{exmp1}  
\end{figure}

\section{Motivating Example}\label{motivex}

Let $\mathcal{P}_{alph}=\{A, B, C, D, E, F, G, H\}$ be a collection of movie IDs. Entities $E_1$ and $E_2$ each have a subset of these IDs as its feasible set as shown in Fig.~\ref{exmp1} with
\begin{align}
    &\mathcal{P}_1=\{A, C, D, G\}, && \mathcal{P}_2=\{B, C, D, G, H\} \nonumber\\
    \implies &X_1 = [1 \, 0 \, 1 \, 1 \, 0 \, 0 \, 1 \, 0]^T, && X_2 = [0 \, 1 \, 1 \, 1 \, 0 \, 0 \, 1 \, 1]^T.
\end{align}
A universal rating system provides a score between $1$ to $5$ to each movie, with $5$ being the best. The goal of the entities is to find the set of movies with the \emph{highest} score present in the list (set) of both entities. Let the scores be,
\begin{align}
    &f(B)=f(E)=f(H)=5, \nonumber\\
    &f(A)=f(C)=f(G)=4 \nonumber\\
    &f(D)=3, \, f(F)=2.
\end{align} 
We have $\bm{\alpha} = [3,1]$. For this problem, $\mathcal{P}^*$ is $\{C,G\}$.

First, the client $E_1$ checks for the presence of $\{A\}$, $\{C\}$ or $\{G\}$ (elements that attain the score of $4$) in $\mathcal{P}_2$, i.e., for indices $\{1,3,7\}$. To do so, $E_1$ picks a random vector ${h}_1$ from $\mathbb{F}_q^8$ with $q>3$ prime and sends the queries, 
\begin{align}
    Q_1 = {h}_1, && Q_2 = {h}_1 + {e}_1 + {e}_3 + {e}_7.
\end{align}
to the first two databases of $E_2$. We assume that the databases of $E_2$ share a common random variable $S_1\in \mathbb{F}_q^8$. The answers returned by the corresponding databases are:,
\begin{align}
    A_1 = Q_1^TX_2 + S_1, && A_2 = Q_2^TX_2 +S_1.
\end{align}
$E_1$ computes $A_2-A_1=2$ and concludes that there are $2$ elements with score $4$ in $\mathcal{P}_1\cap \mathcal{P}_2$. In the first phase, $E_1$ downloads $2$ symbols, one from each database. 

Now, to find which elements out of $\{A,C,G\}$ form the solution set, $E_1$ checks for the presence of any two $\{A,C\}$ (or $\{C,G\}$ or $\{A,G\}$) in $\mathcal{P}_2$. Based on the number of databases $N_2$ at $E_2$'s side, the queries will differ for this phase as tabulated in Table~\ref{tabex}. For instance, if $N_2=3$, the answers downloaded by $E_1$ are $({h}_1+{e}_1)^TX_2+S_1$, from database 3,  ${h}_2^TX_2+S_2$ from database 1 and  $({h}_2+{e}_3)^TX_2+S_2$ from database 2. In this phase, the number of downloaded symbols is $4$, $3$ and $2$ if $N_2$ is $2$, $3$ or $4$, respectively. If there are more than $4$ databases then queries in the last row are sent to any two of databases 3 to $N_2$.  Note that, the download cost in the naive approach for this example is constant at $\ceil{\frac{5N_2}{N_2-1}}$

\begin{table}[t]
\caption{Queries sent to check the presence of $\{A,C\}$ in $\mathcal{P}_2$.}
    \begin{tabular}{|c|c|c|c|c|}
    \hline
      $N_2$ & DB 1 & DB 2 & DB 3 & DB 4\\
      \hline
      2 & ${h}_2$, ${h}_3$ & ${h}_2+{e}_1$, ${h}_3+{e}_3$ & - & -\\
      \hline
      3 & ${h}_2$ & ${h}_2+{e}_3$ & ${h}_1+{e}_1$ & -\\
      \hline
      4 & & & ${h}_1 + {e}_1$ & ${h}_1+{e}_3$\\
      \hline
    \end{tabular}   
    \label{tabex}
\end{table}

Now, with the same feasible sets consider a different score mapping where $f(C)=f(G)=3$ and the rest of the scores are the same as before, $\bm{\alpha}=[1, 3]$ . $E_1$ first checks for the existence of $\{A\}$ in $E_2$ by sending queries ${h}_1$ and ${h}_1+{e}_1$ to two databases with $E_2$. Since the answers from $E_2$ reveal the absence of $A$ in $\mathcal{P}_2$, $E_1$ moves to the next-best value, $3$ and checks for the existence of $\{C\}$, $\{D\}$ or $\{G\}$. On learning that all three elements are common in $\mathcal{P}_2$, $E_1$ directly learns that the optimal solution is $\{C,D,G\}$. The download cost in this scenario is $4$ if $N_2=2$ and $2$ if $N_2>3$. 

Next, consider yet another score mapping where $f(G)=5, f(x)=4, \forall x\in \mathcal{P}_{alph}\setminus \{G\}$, then $\bm{\alpha}=[1, 3]$. This time, $E_1$ starts by checking the existence of $G$ by sending queries ${h}_1$ and ${h}_1+{e}_7$ to two databases with $E_2$. By downloading $2$ symbols as answer, $E_1$ finds the optimal solution to be $\{G\}$.

\section{Achievable Scheme}\label{scheme}

Our achievable scheme to seek for the optimal solution set is an iterative SPIR approach between the client and server where the indices for SPIR are dictated by the function values at the client. The answers received in the current iteration decides the queries in the next. This can also be viewed as a successive implementation of the ThPSI scheme with threshold $t=1$  scheme till a subset that returns positive cardinality is found. Let $M_r:=X_{\mathcal{J}_r}^TX_2$ denote the multiplicity of elements with function value $f_{i_r}$ in $\mathcal{P}_1\cap \mathcal{P}_2$. Further, both the entities agree on a finite field $\mathbb{F}_q$ with $q$ prime, $q > \max_{r\in[L]}\alpha_r$. First, we describe the method CarPSI for any $r\in [L]$.

\subsubsection{CarPSI($X_{\mathcal{J}_r}, N_2$)}  

To compute the cardinality $M_r$, $E_1$ needs to evaluate the dot product of $X_{\mathcal{J}_r}$ with $X_2$. There can be two cases. i) If $r=(k-1)(N_2-1)+1, k\in\mathbb{N}$, $E_1$  uniformly  selects $K$ elements $h_k(j), j\in[K]$ independently from $\mathbb{F}_q$ to form the vector ${h}_k =[h_k(1), h_k(2), \ldots, h_k(K)]$.  Now, $E_1$ randomly selects 2 databases (say the first 2) from the $N_2$ databases available to $E_2$ to participate in the protocol. The queries sent are,
\begin{align}
    Q_1 &= {h}_k && Q_2 = {h}_k+ X_{\mathcal{J}_r}.\label{query2}
\end{align}
The databases at $E_2$ use a new symbol $S_k$ from $\mathcal{S}$. To generate answers, the databases calculate the dot product $Q_n^TX_2$, $n=1,2$ and add the resulting symbol to $S_k$ i.e.,
\begin{align}
    A_1 &= Q_1^T X_2 +S_k=  \sum_{j=1}^K h_k(j) X_2(j) + S_k.\\
    A_2 &= Q_2^T X_2 +S_k= \sum_{j=1}^K \big[h_k(j) X_2(j) + X_{\mathcal{J}_r}(j)X_2(j)\big] +S_k.
\end{align}
All operations are performed in $\mathbb{F}_q$. ii) If $r=(k-1)(N_2-1)+l$, where $l\in \{2,\ldots,N_2-1\}$, then the same vector $h_k$ used in CarPSI($X_{\mathcal{J}_{r-1}}, N_2$) is reused to form the query ${h}_k+ X_{\mathcal{J}_r}$. It is sent to one of the databases $n$ for which $h_k$ was not involved in the query. In response, database $n$ sends $Q_n^T X_2 +S_k$. Having received the answers, $E_1$ computes $M_r$ as,
\begin{align}
    M_r = A_n - A_1 = (Q_n - Q_1)^T X_2 = X_{\mathcal{J}_r}^T X_2.
\end{align}

\begin{algorithm}[t]
\caption{Scheme to find $\mathcal{P}^*$}\label{alg1}
 \hspace*{\algorithmicindent} \textbf{Input:} $\mathcal{P}_1,f,\bm{\alpha},N_2$\\
 \hspace*{\algorithmicindent} \textbf{Output:} $\mathcal{P}^*$
\begin{algorithmic}
\State $L \gets \hbox{ length of } \bm{\alpha}$
\For {$r \in \{1,2,\ldots, L\}$}
\If {$r<L$ or $\alpha_L>1$}
\State $M_r = \text{CarPSI}(X_{\mathcal{J}_r},N_2)$
\If{$1\leq M_r\leq \alpha_r-1$}
\State $\mathcal{P}^*$ = FindPSI($\mathcal{J}_{r},M_r, N_2$)
\State \hspace*{\algorithmicindent} \textbf{break}
\ElsIf {$M_r==\alpha_r$}
\State $\mathcal{P}^*=\mathcal{P}_{alph}(\mathcal{J}_r)$
\State \hspace*{\algorithmicindent} \textbf{break}
\EndIf
\Else 
\State $\mathcal{P}^*=\mathcal{P}_{alph}(\mathcal{J}_r)$
\EndIf
\EndFor
\State \Return $\mathcal{P}^*$
\end{algorithmic}
\end{algorithm}

The complete flow of the achievable scheme is given in Algorithm~\ref{alg1}. First, $E_1$ creates the sets of indices $\mathcal{J}_r, r\in[L]$ and the ordered vector $\bm{\alpha}$ of length $L$. Next, it executes CarPSI($X_{\mathcal{J}_1},N_2$) to find $M_1$ for the best function value $f_{i_1}$. Based on $M_1$ and $\alpha_1$, there are 3 cases:
\begin{itemize}
    \item [i)] If $M_1=0$, $E_2$ does not possess any of the elements in common with $E_1$ that yield the same function value $f_{i_1}$. $E_1$ proceeds to the next best function value $f_{i_2}$ and finds $M_2$ by executing CarPSI($X_{\mathcal{J}_2},N_2$). 
    \item [ii)] If $M_1=\alpha_1$, then $f_{i_1}$ is a feasible function value with $\mathcal{P}_{alph}(\mathcal{J}_1)$ as the solution set. 
    \item [iii)] If $1\leq M_1\leq \alpha_1 -1$, $E_1$ proceeds to find the solution set which comprises $M_1$ out of $\alpha_1$ indices in $\mathcal{J}_1$ by implementing FindPSI($\mathcal{J}_1,M_1,N_2$).
\end{itemize} 

In general,  CarPSI is executed till the returned cardinality $M_r\geq 1$, $r\in [L]$. If $\alpha_r=M_r$, $E_1$ directly sets $\mathcal{P}^*=\mathcal{P}_{alph}(\mathcal{J}_r)$. Otherwise, $E_1$ executes FindPSI once. If the search continues and $M_r=0$ for every $r\leq L-1$, then the solution is contained in $\mathcal{P}_{alph}(\mathcal{J}_L)$ since we assumed that $|\mathcal{P}_1\cap \mathcal{P}_2| \geq 1$. 

\subsubsection{FindPSI($\mathcal{J}_r,M_r,N_2$)} 

Since $M_r$ is known, it suffices if $E_1$ leaves any single index (say, $j$) from $\mathcal{J}_r$ out, to learn the set intersection. Let $\mathcal{\Bar{J}}=\mathcal{J}_r\setminus \{j\}$ be the set on which PSI is to be executed.  Note that $|\Bar{\mathcal{J}}|=\alpha_r-1$. Depending on the value of $r$, two cases arise. i) If $r=k(N_2-1)+l$, with $k\in\mathbb{N}\cup\{0\}, l\in\{1,\ldots, N_2-2\}$, this means that $l+1$ databases have been previously sent queries to using the random vector $h_k$. $E_1$ reuses ${h}_k$ to submit queries to the unused (assume, $l+2,\ldots,N_2$) databases. Similarly, the databases reuse the randomness symbol $S_{k}$. ii) If $r= k(N_2-1),$ $k\in \mathbb{N}$, then $E_1$ generates a fresh random vector ${h}_{k+1}\in \mathbb{F}_q^K$ and $E_2$ picks a new randomness symbol $S_{k+1}$ from $\mathcal{S}$ to start FindPSI.  

We write $\alpha_r-1=(N_2-l-1)+ p(N_2-1) + s$, $p\in \mathbb{N}\cup \{0\}, s\in [N_2-2]$. For $k<l$, we use the notation ${k:l}$ to denote the set of vectors $\{k, k+1,\ldots,l\}$. The queries sent to the databases and the answers returned are,
\begin{align}
Q_n&=
\begin{cases}
h_y, & \text{ for } n=1 \\
{h}_y+ {e}_j, j\in \Bar{\mathcal{J}} & \text{ for }n\in[N_2]\setminus\{1\},
\end{cases} 
\end{align}
and
\begin{align}
A_n&=Q_n^TX_2 +S_y, \quad n\in[N_2].
\end{align} 
In case ii) $y\in {k+1:k+p+1}$ if $n\in[s+1]$ and $y\in {k+1:k+p}$ otherwise. In case i) if $s\leq l$, then $y\in {k+1:k+p+1}$ if $n\in[s+1]$, $y\in {k+1:k+p}$ if $n\in s+2:l+1$ and $y\in {k:k+p}$ if $n\in {l+2:N_2}$. The sub-case where $s>l$ can be similarly handled.

To decode each element in $X_2(\Bar{\mathcal{J}})$, $E_1$ evaluates $A_n-A_1$. By doing so, $E_1$ learns $X_2(\mathcal{J}_r)$ since $M_r$ is known. 

\begin{remark}
    Since each answer is secured by a unique common randomness symbol, $S_k\in \mathcal{S}$, it is sufficient to have $|\mathcal{S}|=\ceil{\frac{R+\alpha_R}{N_2-1}}$ for the feasibility of our scheme. However, $R,\alpha_R$ being unknown to $E_2$ before communication, the databases share $\ceil{\frac{P_1}{N_2-1}}$ common randomness symbols.
\end{remark}

\begin{remark}
    If $M_r = X_{\mathcal{J}_r}^T X_2=0$ $\forall r\in[L-1]$ and $\alpha_L=1$, then $|\mathcal{P}_1\cap \mathcal{P}_2|=1$ and the only element in $\mathcal{P}_1$ whose availability in $\mathcal{P}_2$ is not checked, is the solution. Thus, $R=L, \alpha_R=1$ and the download cost is $D_{PSI}(L-1,N_2)=\big\lceil\frac{(L-1)N_2}{N_2-1}\big\rceil<\big\lceil\frac{P_1N_2}{N_2-1}\big\rceil$ since the $L^{th}$ round of CarPSI and FindPSI are skipped.
\end{remark}

\subsubsection{Correctness and Privacy}
Correctness of the achievable algorithm directly follows from the correctness of SPIR scheme. $E_1$'s privacy with respect to each database of $E_2$ is guaranteed, because the query vectors received by databases with $E_2$ are random with each element appearing to be chosen uniformly from $\mathbb{F}_q$. Protection against decoding any additional information from the successive queries is also guaranteed, since every new query sent to a database is generated with a fresh random vector $h_k$ which is private to $E_1$. The privacy of $E_2$ is guaranteed because every new answer sent by a database is added to a randomness symbol $S_k\in \mathcal{S}$, private to $E_2$.

\begin{remark}
    $E_1$ learns that, the elements in $\mathcal{P}_1 $ corresponding to indices $\cup_{r=1}^{R-1}\mathcal{J}_r$ are absent in $E_2$'s constraint set, $\mathcal{P}_2$,  ($X_2$ is 0 at those indices) and the value of $X_2(\mathcal{J}_R)$. Thus, $E_1$ learns $X_2$ at $\cup_{r=1}^{R}\mathcal{J}_r \subset \mathcal{I}_1$ and
\begin{align}
I(&\mathcal{P}_2;Q_{1:N_2},A_{1:N_2},\mathcal{P}_1,f) \nonumber\\
&= H(\mathcal{P}_2) -   H(\mathcal{P}_2|Q_{1:N_2},A_{1:N_2},\mathcal{P}_1,f) \\
&= H(X_2)-H\big(X_2([K] \setminus \cup_{r=1}^{R}\mathcal{J}_r)\mid X_2(\cup_{r=1}^{R}\mathcal{J}_r)\big)\\
&=H\big(X_2(\cup_{r=1}^{R}\mathcal{J}_r)\big), \label{infoleak}
\end{align}
where we used the fact that $X_i$ is a sufficient statistic for $\mathcal{P}_i$. In fact, \eqref{infoleak} is equal to \eqref{nominfoleak}. 
\end{remark}
 
\subsubsection{Proof of Theorem~\ref{thm1}}
We have that $R$ is the least value of $r\in[L]$ in Algorithm~\ref{alg1} at which $M_r\geq 1$.  The download cost for the sequential CarPSI is equal to $D_{PSI}(R,N_2)$ which is $\big\lceil\frac{RN_2}{N_2-1}\big\rceil$. This is the download cost if $M_R=\alpha_R$. 

Otherwise, let  $R=k(N_2-1)+l$, $k\in\mathbb{N}\cup \{0\}$. If $l=0$, the download cost in FindPSI is $\big\lceil\frac{(\alpha_R-1)N_2}{N_2-1}\big\rceil$ and that from CarPSI is $\frac{RN_2}{N_2-1}$. Their sum gives us \eqref{dlcost1}. 
With $l\in[N_2-2]$, after CarPSI($X_{\mathcal{J}_R}, N_2$), $l+1$ out of $N_2$ databases  are used. The remaining $N_2-(l+1)$ databases continue to send answers for FindPSI using the same random vector for query and common randomness symbols respectively, resulting in the partial download cost,
\begin{align}
    &\bigg\lceil{\frac{RN_2}{N_2-1}}\bigg\rceil+N_2-l-1= (k+1)N_2. \label{dpart}
\end{align}
This is followed by downloads for $\alpha_R-1-(N_2-l-1)$ indices which costs $D_{PSI}(\alpha_R-N_2+l,N_2)$, which when added to \eqref{dpart} yields,
 \begin{align}
    D &= (k+1)N_2 + \bigg\lceil\frac{(\alpha_R-N_2+l)N_2}{N_2-1}\bigg\rceil\\
     &=\bigg\lceil\frac{(\alpha_R-1)N_2+(k(N_2-1)+l)N_2}{N_2-1}\bigg\rceil\\
     &= \bigg\lceil\frac{(\alpha_R-1+R)N_2}{N_2-1}\bigg\rceil.\label{lst}
 \end{align}
 
\subsubsection{Proof of Theorem~\ref{thPSI}}
The achievable scheme of ThPSI first finds the cardinality of set intersection, $M = |\mathcal{P}_1\cap \mathcal{P}_2|$. If $M$ is at least the value of a threshold $t\in\mathbb{N}$, the set intersection is found obeying the privacy constraints of PSI. 
Under the same system model as described in Section~\ref{mod}, the client $E_1$ computes $M=X_1^TX_2$ by executing CarPSI($X_1,N_2$) with $\mathbb{F}_q$ having $q>P_1$, prime. If $t\leq M\leq P_1-1$, then it executes FindPSI($\mathcal{I}_1,M,N_2)$. Depending on $M$, there are 2 cases:
\begin{itemize}
    \item[i)] $M<t$ or $M=P_1$:  $D_{ThPSI}=2$ from CarPSI.
    \item [ii)] $t\leq M\leq P_1-1$: On fulfilling the threshold condition,
    \begin{align}
    \!\!\!D_{ThPSI} =& \underbrace{2}_{\hbox{CarPSI}} \nonumber \\
    &+ \underbrace{(N_2 -2) + D_{PSI}(P_1-N_2+1,N_2)}_{\hbox{FindPSI}}\\
    =& N_2 + \bigg\lceil\frac{(P_1-N_2+1)N_2}{N_2-1}\bigg\rceil \\
    =& \bigg\lceil\frac{N_2(N_2-1) +P_1 N_2 -N_2(N_2-1)}{N_2-1}\bigg\rceil  \\
    =& \bigg\lceil\frac{P_1 N_2}{N_2-1}\bigg\rceil.
\end{align}
\end{itemize}

\subsubsection{Proof of Remark~\ref{remPeq}}
With a fixed realization of $\mathcal{P}_1$ and $\mathcal{P}_2$, the random variable $R\in [L]$ and vector $\bm{\alpha}$ depend on the realization of $f$. The maximum value of $L$ (hence $R$) is $R_{\max}=\min(T,P_1-M+1)$. Further, for any $R\in [L]$, $M_R\leq\alpha_R$. Our scheme's download cost reduces to that of the naive scheme when the cumulative download costs from $R$ rounds of CarPSI and one round of FindPSI is equal to $D_{PSI}$. Additionally, $M_R$ must be strictly less than $\alpha_R$ so that FindPSI is executed. Thus, $D=D_{PSI}$ if $\{R+\alpha_R=P_1+1\}$ and $\{M_R<\alpha_R\}$ are simultaneously satisfied. Note that, for such events $\alpha_r=1$ for all $r<R$. Further, $M_r=0$ for all $r<R$, hence the function value of every element in $\mathcal{P}_1\cap \mathcal{P}_2$ is $f_{i_R}$. With $M \geq 1$, the corresponding probability $P_{eq}$ is,
\begin{align}
    P_{eq}&=\mathbb{P}(\{R+\alpha_R=P_1+1, M_R<\alpha_R\})\\
    &=\sum_{r=1}^{R_{\max}}\mathbb{P}(R=r,\alpha_r=P_1+1-r, M_r<P_1+1-r) \label{Mralphar} \\
    &=\sum_{r=1}^{R_{\max}}\mathbb{P}(R=r,\alpha_r=P_1+1-r)\mathbb{P}(M<P_1+1-r) \label{Malphar}\\ 
    &=\sum_{r=1}^{R_{\max}}P_{eq}(r)\times\mathbbm{1}_{\{M<P_1-r+1\}}\label{indic}.
\end{align}
where \eqref{Malphar} follows from \eqref{Mralphar} using $\mathbb{P}(M_r<P_1+1-r|R=r,\alpha_r=P_1+1-r)=\mathbb{P}(M<P_1+1-r)$ since $M_R=M$. Further, since $M$ is determined by $\mathcal{P}_1$ and $\mathcal{P}_2$, $\mathbb{P}(M<P_1+1-r)$ is an indicator as in \eqref{indic}. If $r=1$, we have
\begin{align}
   P_{eq}(1)=T\bigg(\frac{1}{T}\bigg)^{P_1},
\end{align}
since $T$ out of $T^{P_1}$ function realizations of assume equal function value over $\mathcal{P}_1$. For $1<r\leq R_{\max}$, 
\begin{align}
    \label{peqr}
    P_{eq}(r)=&\binom{P_1-M}{r-1}(r-1)!\bigg[\sum_{j=r-1}^{T-1}\binom{j-1}{r-2}\times(T-j)\bigg]\nonumber\\
    &\times\bigg(\frac{1}{T}\bigg)^{P_1},
\end{align}
where the terms outside the summation in \eqref{peqr} follow by sampling $f_{i_1},f_{i_2},\ldots,f_{i_{r-1}}$ with arrangements from the $P_1-M$ elements in $\mathcal{P}_1$ absent in $\mathcal{P}_1\cap \mathcal{P}_2$, multiplied by the probability of each mapping. Fixing $f_{i_{r-1}}=j$, we choose $f_{i_1},\ldots,f_{i_{r-2}}$ from $1$ to $j-1$ as better function values preceding $f_{i_{r-1}}$, while the remaining $\alpha_R=P_1-r+1$ elements have $T-j$ choices for $f_{i_R}$ from $\{j+1,j+2,\ldots,T-j\}$. Finally, we sum over all  $j\in\{r-1,\ldots,T-1\}$.

\begin{figure}[t]
   \centering
   \includegraphics[width=0.45\textwidth]{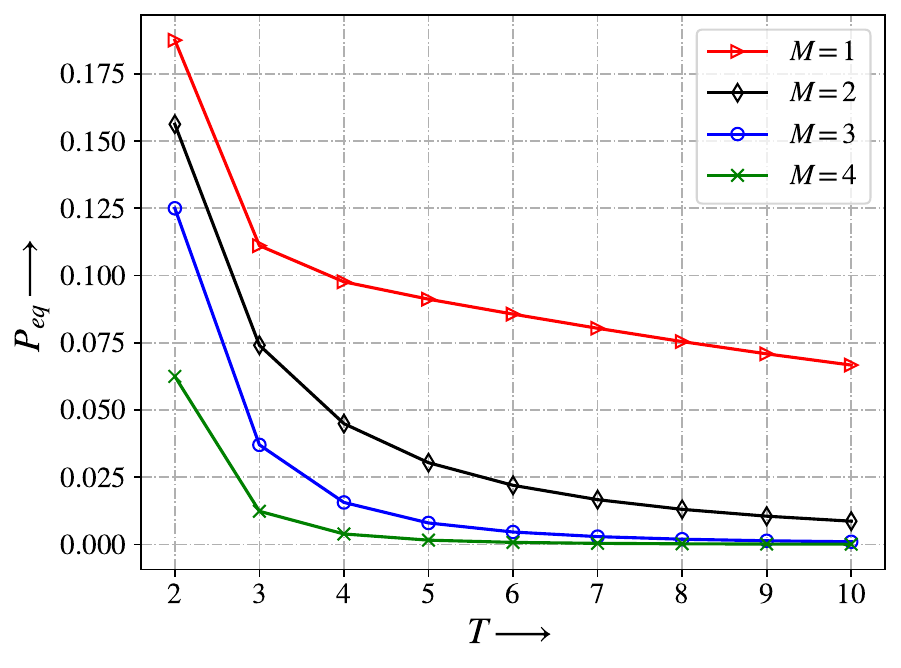}
   \caption{$P_{eq}$ against $T$ for different values of $M$.}
   \label{probTM}
\end{figure}

To show that \eqref{indic} is small, in Fig.~\ref{probTM}, we plot the probability $P_{eq}$ with respect to the size of function range $T$ for various values of $M$. We follow the example of movie scores in Section~\ref{motivex} and fix $P_1=5$ while the scores are allotted from the set $\mathcal{T}=[T]$ with $T\in\{2,3,\ldots,10\}$. For a fixed $T$, increasing $M$ from $1$ to $4$ decreases the value of $P_{eq}$. 

\section{Conclusion}\label{conclude}

We proposed a two-agent function optimization algorithm under information theoretic privacy of feasible sets. Our algorithm runs the primitives CarPSI and FindPSI, built on optimal SPIR schemes to find the optimum solution set. In doing so, the information on the feasible set of an agent leaked to the other agent is kept at a minimum. We showed that both the download cost and the information leakage of our scheme are lower than those of an alternative (naive) scheme that relies on learning the joint feasible set using PSI. As a byproduct of our algorithm, we characterized the worst case download cost of an information theoretically private ThPSI protocol. 

\bibliographystyle{unsrt}
\bibliography{refs}
\end{document}